\documentclass[aps,prb,twocolumn,showpacs,floatfix]{revtex4}
\usepackage{graphicx}
\usepackage[centertags]{amsmath}

\newcommand{\icm}{\ensuremath{\;\mbox{cm}^{-1}}}
\newcommand{\K}{\ensuremath{\;\mbox{K}}}
\newcommand{\celsius}{\ensuremath{^\circ}\!C}

\begin{document}

\title{Far-infrared soft mode behavior in
$\mathbf{PbSc_{1/2}Ta_{1/2}O_3}$ thin films}

\author{S.~Kamba, M.~Berta, M.~Kempa, and J.~Petzelt}
\affiliation{Institute of Physics, Academy of Sciences of the
Czech Republic, Na Slovance~2, 182 21 Prague~8, Czech Republic}
\author{K.~Brinkman and N.~Setter}
\affiliation{Ceramics Laboratory, EPFL, Swiss Federal Institute of
Technology, 1015 Lausanne, Switzerland}

\date{\today}

\pacs{78.30.-j; 63.20.-e;77.22.-d; 77.84Dy}

\begin{abstract}
Temperature dependences of the optic phonons in $\rm
PbSc_{1/2}Ta_{1/2}O_3$ sol-gel films deposited on sapphire
substrates were studied by means of Fourier transform far-infrared
transmission spectroscopy in the temperature range 20-900\K. Four
films displaying different B-site order with both ferroelectric
and relaxor behavior were studied. In all cases the TO mode near
80\icm{} at 10\K{} softens on heating to $\approx$45\icm{}
following the Cochran law with extrapolated critical temperature
near 700\K{} (400\K{} above the temperature of dielectric maximum,
$T_{\rm m}$), but above 600\K{} its frequency remains stabilized.
It can be assigned to the A$_{1}$ component of the ferroelectric
soft mode inside polar clusters which form below the Burns
temperature near 700\K. In the ordered PST film another mode
activates below T$_{m}$ in infrared spectra near 60\icm\
exhibiting also anomalous temperature dependence due to its
coupling with the former mode. It is assigned to the A$_{1}$
component of the F$_{2g}$ Raman active mode. Central mode, which
appears below the Burns temperature in the THz range, is assigned
to the dynamics of polar clusters. It slows down on cooling and
vanishes from our spectral range below $T_{\rm m}$. Another
overdamped excitation assigned to the E component of the soft mode
appears near 30\icm{} at low temperatures.
\end{abstract}

\maketitle

\section{Introduction}

Ferroelectric relaxors, in particular complex perovskites with the general formula $\rm
PbB_{1/3}B''_{2/3}O_3$ and $\rm PbB'_{1/2}B''_{1/2}O_3$, are of high interest due to
their excellent dielectric, electrostrictive, and pyroelectric properties\cite{Samara01}.
A discovery of giant piezoelectric response in relaxor based crystals by Park and
Shrout\cite{Park97} turned attention of scientific community to better understanding the
relaxor ferroelectricity.

Lead magnesium niobate $\rm PbMg_{1/3}Nb_{2/3}O_3$ (PMN) and lead
scandium tantalate $\rm PbSc_{1/2}Ta_{1/2}O_3$ (PST) are model
representatives of the relaxor ferroelectrics which exhibit high
and broad maxima in the real $\varepsilon'(T)$ and imaginary
$\varepsilon''(T)$ part of the dielectric permittivity which shift
with the increasing frequency to higher temperatures. No
ferroelectric (FE) phase transition occurs in PMN without bias
electric field, while PST exhibits a~spontaneous FE transition.
Its temperature depends on the ordering of B-site ions (Sc and
Ta). Disordered PST undergoes FE transition at $T_{\rm
c}\sim270\;$K, while $T_{\rm c}$ of the ordered sample appears
near 300\K{} without typical relaxor behavior above $T_{\rm
c}$\cite{Setter80,Chu93}.

It is well known that the peculiar dielectric properties of
relaxors are connected with a~broad dielectric relaxation (we call
it central mode (CM) in analogy with inelastic scattering
experiments) below the polar phonon frequencies. It stems from the
dynamics of nanoscopic inhomogeneities - polar clusters. Many
low-frequency experimental dielectric data exist on the behavior
of CM in FE relaxors, but much less data were published about
dielectric response in the MHz and GHz
range\cite{Kamba04a,Buixaderas04} and THz data are practically
missing in the literature. Neutron diffuse scattering in PMN
revealed\cite{Hirota02} the CM below the Burns temperature,
$T_{\rm d}=620\;$K, where the polar nanoclusters appear. However,
its frequency was not determined.

Very recently, infrared (IR) transmission and microwave dielectric
data\cite{Kamba05} of PMN have shown that the CM has
characteristic frequency near 20\icm\ (0.6\,THz) at $T_{\rm d}$
and its frequency dramatically slows down to sub-Hertz region on
cooling to freezing temperature $T_{\rm f}$ near
200\K\cite{Bovtun04}. Simultaneously the relaxation broadens on
cooling, giving rise to frequency independent losses between
100$\;$Hz and 100$\;$GHz at temperatures below $T_{\rm
f}$\cite{Bovtun04}. Dielectric studies of PST up to 33$\;$GHz
revealed a~two-component CM,\cite{Chu96,Bovtun00} however its
behavior at high temperatures near $T_{\rm d}$ was not studied.
One of the aims of this study is to investigate the CM at high
temperatures to confirm its vanishing (merging with the soft mode
response) above $T_{\rm d}$.

Lattice dynamics of PMN was studied by means of Raman, IR and inelastic neutron
scattering spectroscopy\cite{Kamba05,Bovtun04,Siny00,Wakimoto02b,Shirane03}. It was
shown\cite{Wakimoto02b,Kamba05} that the lowest frequency transverse optic phonon
(TO$_1$) $\omega_{\rm SM}$ follows the Cochran law
\begin{equation}
  \label{eq:Cochran}
  \omega_{\rm SM}=A\sqrt{T_{\rm d}-T}
\end{equation}
with the extrapolated critical temperature near the Burns
temperature $T_{\rm d}=620\;$K. It indicates that the TO$_1$
frequency $\omega_{\rm SM}$ can be assigned to the FE soft mode
(SM) in polar clusters. The questions arise: Is it the general
behavior typical for all FE relaxors? What is the behavior of
relaxors with the spontaneous FE phase transition (PT)? Does the
soft mode feel the FE PT temperature $T_{\rm c}$ (like in other
displacive FEs) or rather the Burns temperature which is typically
300-400\K\ above $T_{\rm c}$?

The phonons in bulk PST were already studied by
Raman\cite{Setter87,Bismayer89,Siny89,Mihailova02,Guttler03} and
IR spectroscopy\cite{Reaney94,Petzelt98} and no optic mode
softening was observed. Only small anomalies in phonon frequencies
were seen in the Raman and IR spectra near $T_{\rm c}$. From that
it was concluded that the transition is of the order-disorder
type, which is, however, rather unusual for perovskite FEs.

Also the question of phonon activity in the Raman spectra is of
interest. Factor-group analysis for the Pm3m space group in the
paraelectric phase, corresponding to disordered PST, predicts
three F$_{\rm 1u}$ only IR active phonons and one silent F$_{\rm
2u}$ mode so that no phonons should be Raman active. Nevertheless,
a number of modes were observed in the Raman spectra of
paraelectric phase. Siny \textit{et al.}\cite{Siny89} and
independently Bismayer \textit{et al.}\cite{Bismayer89} explained
the activity of the phonons in Raman spectra by doubling of the
unit cell in the paraelectric phase due to the 1:1 B-site
ordering. In this case the space group is Fm$\overline{3}$m and
the factor group analysis yields the following optic vibration
modes:
\begin{multline}
  \label{factor}
  \mathrm{\Gamma = A_{\rm 1g}(R) + E_{g}(R) + F_{1g}(-) +}\\
  \mathrm{+ 4F_{1u}(IR) + 2F_{2g}(R) + F_{2u}(-).}
\end{multline}

It means that 4~Raman and distinct 4~IR active modes should be
expected in the spectra. However, first-order Raman lines are seen
not only in B-sites ordered samples, but also in the disordered
PST and in other relaxors like PMN. Siny \textit{et
al.}\cite{Siny89} explained this Raman activity by local
short-range B-site ordering in chemical clusters. Namely, Raman
scattering is sensitive to short-range (nm) order, as the mode
activity is given by the phonon eigenvectors which are determined
by the interatomic force constants.

Another assignment was suggested recently by Hlinka \textit{et
al.}\cite{Hlinka05}. They argued that Raman scattering in relaxors
is primarily due to anisotropic polar clusters rather than to 1:1
B-site ordering. It would mean that the Raman-activated modes
belong prevailingly to the Brillouin-zone center of the parent
Pm3m cubic structure rather than to modes activated by the zone
folding as it was assumed by Siny \textit{et al.}

Also results of the pulsed neutron atomic pair-density function
analysis by Egami \textit{et al.}\cite{Egami03} speak in favor of
such a picture. They have shown that the Burns temperature $T_{\rm
d}$ is the local Curie temperature below which the polar clusters
are formed, but the local polarization of Pb persists up to
$\sim1000\;$K several hundreds of Kelvins above $T_{\rm d}$.
Probably both mechanisms of mode activation in Raman spectra have
to be considered and to distinguish between them requires more
detailed studies, particularly at very high temperatures near the
B-site ordering temperature. Comparison of the Raman spectra with
properly evaluated IR spectra is also necessary.

Factor-group analysis in the R3m rhombohedral ferroelectric phase
with the doubled unit cell due to B-site ordering yields the
following optic modes:
\begin{equation}
  \label{factor 2}
  \rm  \Gamma = 7A_{1}(R,IR) + 2A_{2}(-) + 9E(R,IR),
\end{equation}
and $\rm 1A_{1} + 1E$ acoustic modes. It means that up to 16 TO
modes simultaneously Raman and IR active plus corresponding 16 LO
Raman active modes can be expected in the spectra of the
ferroelectric phase. IR active modes in bulk PST ceramics were
studied by means of IR reflectivity\cite{Reaney94,Petzelt98}. The
spectra revealed correlation between the degree of 1:1 B-site
ordering and appearance of an extra mode at 315\icm. The
high-frequency wing of the CM was observed above room temperature
in the reflectivity spectra below 30\icm. However, the accuracy of
IR reflectivity technique is limited at low frequencies and high
temperatures. In this case transmission measurements are more
accurate and more sensitive to absorption mechanisms, but for this
purpose, due to high absorption, thin films are needed. The aim of
this study is to investigate temperature dependence of the
low-frequency polar phonons and CM in PST films with various
B-site order in a broad temperature interval 20 - 900\K. It should
allow us to shed more light on the lattice and polar-cluster
dynamics between $T_{\rm c}$ and the Burns temperature $T_{\rm
d}$.

\section{Experiment}

PST thin films were prepared by chemical solution deposition on sapphire substrates,
which are transparent in the far IR \cite{Brinkman04}. Four films of various degree of
order were investigated: disordered (annealed at 700\celsius{} for 1$\;$min), slightly
ordered (annealed at 850\celsius{} for 1$\;$min), 50\% ordered (annealed at 850\celsius{}
for 1$\;$hour) and 78\% highly ordered (annealed at 800\celsius{} for 48$\;$hours). The
degree of ordering was determined by the X-ray diffraction from
$(\frac{1}{2},\frac{1}{2},\frac{1}{2})$ superlattice peak intensity and the size of the
ordered regions was investigated using TEM dark-field imaging.\cite{Brinkman04} The
thickness of the films was about 500$\;$nm, plane-parallel (0001) oriented sapphire
substrates were 0.5$\;$mm thick. The unpolarized IR transmission spectra were taken using
FTIR spectrometer Bruker IFS 113v at temperatures between 20 and 900\K{} with resolution
of 0.5\icm. A helium-cooled Si bolometer operating at 1.5\K{} was used as detector, an
Optistat CF cryostat with polyethylene windows was used for cooling, and a
high-temperature cell SPECAC P/N 5850 was used for the heating. The investigated spectral
range was determined by the transparency window of the sapphire substrate; at 20\K{} up
to 450\icm{} (15$\;$THz), at 900\K{} the sample was opaque already above 190\icm.
\begin{figure}[htbp]
  \begin{center}
    \includegraphics[width=\columnwidth]{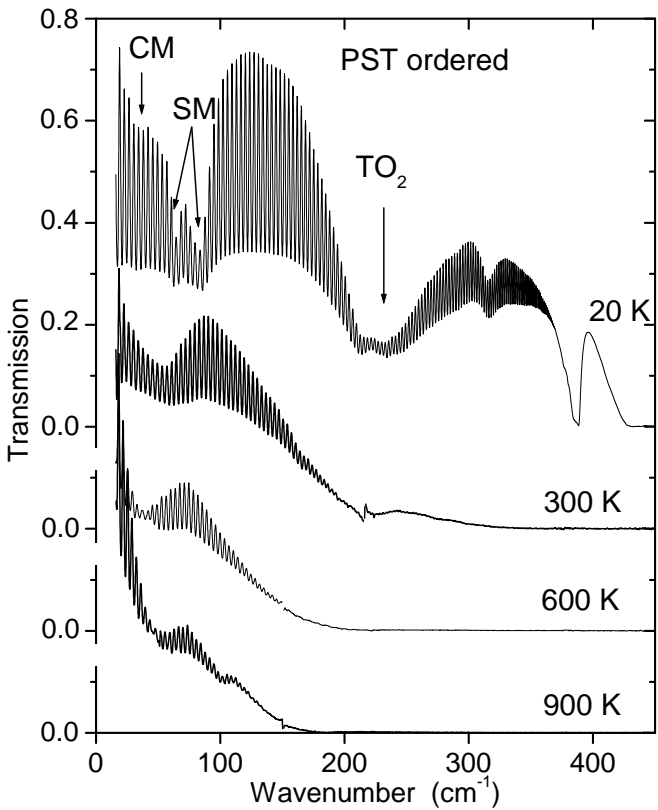}
  \end{center}
  \caption{IR transmission spectra of the 500$\;$nm thick 78\%
  ordered PST film deposited on sapphire substrate (490$\;\mu$m)
  for various temperatures. Frequencies of the central mode (CM),
  split TO$_1$ soft mode (SM) and TO$_2$ phonon are marked.
  Absorption peak near 380\icm{} is a~phonon peak from the
  sapphire. Small peak near 220\icm{} at 300\K\ is an instrumental effect.}
  \label{fig:PSTo_trans}
\end{figure}

\section{Spectra evaluation}

IR transmission spectra of 78\% ordered PST film at selected temperatures are shown in
Fig.~\ref{fig:PSTo_trans}. Dense oscillations in the spectra are due to interferences in
the substrate, while broad minima correspond to frequencies of polar phonons in the film.
Splitting of TO$_1$ SM and TO$_2$ phonons at 20\K{} is clearly seen. The spectra were
taken at more than 20 temperatures. At each temperature the spectra of a~bare sapphire
substrate and the PST film on the substrate were measured. The overall transmission
decreases on heating, mainly due to the increase in multi-phonon absorption of the
sapphire substrate. The transmission spectrum of the bare substrate was first fitted with
a~sum of harmonic oscillators using Fresnel formulae for coherent transmission of
a~plane-parallel sample (i.e. taking into account the interference effects)\cite{Born}.
The resulting fit parameters of sapphire were then used for the fit of the PST/sapphire
two-layer system.

The complex transmittance of the two-layer system was computed by
the transfer matrix formalism method including interference
effects\cite{Heavens}. Model of the sum of damped quasi-harmonic
oscillators in the form
\begin{equation}
  \label{eps3p}
  \varepsilon^*(\omega) = \varepsilon'(\omega)-\textrm{i}
  \varepsilon''(\omega) = \varepsilon_{\infty} + \sum_{j=1}^{n}
  \frac{\Delta\varepsilon_{j}\omega_{j}^{2}} {\omega_{j}^{2}
  - \omega^2+\textrm{i}\omega\gamma_{j}}
\end{equation}
was used for the expression of the complex permittivity
$\varepsilon^*$ of the films. $\omega_{j}$, $\gamma_{j}$ and
$\Delta\varepsilon_{j}$ denote the frequency, damping and
contribution to the static permittivity of the $j$-th polar mode,
respectively. The parameter $\varepsilon_{\infty}$ describes the
high-frequency permittivity originating from the electronic
polarization and from the polar phonons contribution above the
spectral range studied.
\begin{figure}[htbp]
  \begin{center}
    \includegraphics[width=\columnwidth]{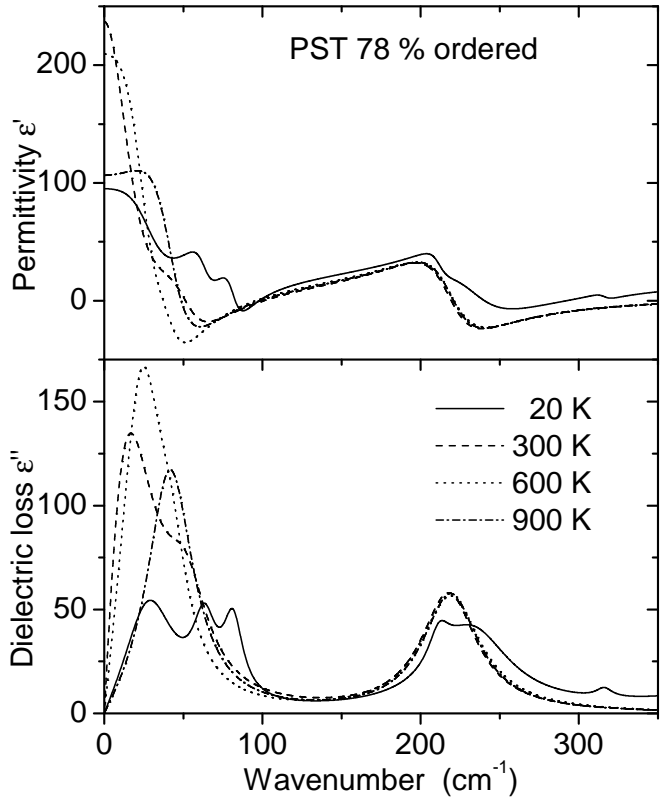}
  \end{center}
  \caption{Dielectric spectra of 78\% ordered PST film obtained
  from the fit of transmission spectra for various temperatures.}
  \label{fig:PSTo_eps}
\end{figure}

\section{Results and discussion}

Real ($\varepsilon'$) and imaginary ($\varepsilon''$) parts of
permittivity $\varepsilon^*$ of the 78\% ordered PST thin film
calculated from the fit of transmission spectra in
Fig.~\ref{fig:PSTo_trans} are shown in Fig.~\ref{fig:PSTo_eps}.
The frequencies of maxima of $\varepsilon''$ correspond to the
mode frequencies. Parameters of the modes above 200\icm{} were
fixed during our fits at temperatures above 300\K\ because the
substrate was opaque and this range is omitted in the figure. We
note that the mode parameters correspond very well to parameters
obtained from the bulk IR reflectivity spectra published in Ref.
\cite{Petzelt98}. It shows that the phonon parameters are not
influenced by possible strain or size effect in our thin films. We
believe that the IR transmission spectra are more sensitive than
the IR reflectivity spectra. It could be the reason why we could
see also the splitting of the TO$_{2}$ mode although it was not
resolved in the IR reflectivity spectra.
\begin{figure}[htbp]
  \begin{center}
    \includegraphics[width=\columnwidth]{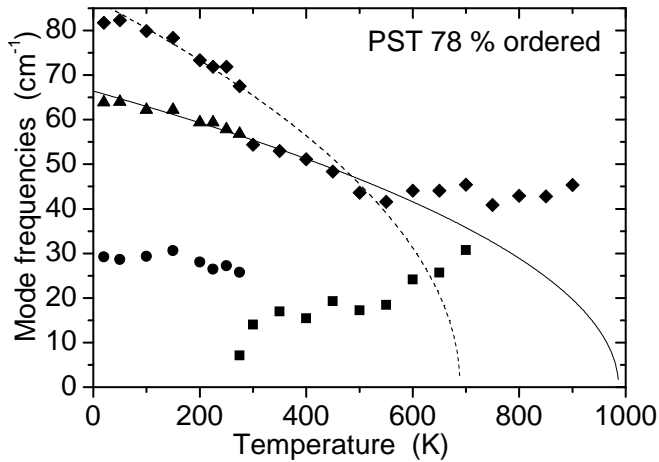}
  \end{center}
  \caption{Temperature dependences of the polar mode frequencies
  in 78\% ordered film. The lowest frequency mode below 30\icm{}
  (marked as CM) is overdamped, therefore the frequency of loss
  maximum corresponding to $\omega_{\rm CM}^2/\gamma_{\rm CM}$ is
  plotted. Two kinds of the Cochran fit of the SM (shown by solid
  and dashed lines) are discussed in the text.}
  \label{fig:PSTo_mod}
\end{figure}

The most important mode anomalies were observed in the range below
100\icm. Temperature dependences of the polar mode frequencies are
plotted in Fig.~\ref{fig:PSTo_mod}. A~new heavily damped (i.\,e.
$\gamma_j/\omega_j>1$) excitation appears on cooling below 700\K{}
at frequencies near 30\icm. This mode is not manifested by a
minimum in the IR transmission spectra, but rather by enhanced
broad absorption at below 40\icm. This excitation does not seem to
be of usual one-phonon origin, and we assign it to dynamics of
polar clusters (i.e. CM).

Disappearance of the CM above 700\K{} is manifested by an enhancement of the
low-frequency transmission, although the high-frequency transmission continues to
decrease on heating, as expected due to increase in damping of polar phonons and
multi-phonon absorption (see Fig.~\ref{fig:PSTo_trans}). On cooling to $T_{\rm c}
\approx$300\K, CM slows down to microwave range and vanishes below $T_{\rm c}$ from our
spectral range.

Another heavily damped excitation arises in our IR spectra below
$T_{\rm c}$ near 30\icm\ and remains in the spectra down to the
lowest temperatures. The origin of this excitation was not fully
understood until recently (in analogy to PMN, where it also
appears, \cite{Kamba05} it could be explained by activation of the
acoustic phonon branches due to the local unit cell doubling or
translation symmetry breaking). However, now it seems to us more
probable that it is just the E component of the split F$_{1u}$ SM
(for this explanation see below).

Let us stress that our static permittivity in
Fig.~\ref{fig:PSTo_eps} does not correspond to experimental
low-frequency permittivity, because strong dielectric dispersion
exists below the phonon frequencies.\cite{Bovtun00} Bovtun
\textit{et al.} combined high-frequency and microwave dielectric
spectra of PST ceramics between 1 MHz and 36 GHz\cite{Bovtun00}
with the IR reflectivity spectra \cite{Petzelt98} and revealed two
dispersion regions below phonons. One, nearly temperature
independent, was estimated to near 1\,THz and corresponds well to
our heavily damped mode near 30\icm\ below T$_{c}$. Second
relaxation was seen at temperatures below 350\K\ in the
high-frequency range and it slows down and broadens on cooling. At
higher temperatures it merges with the far IR spectra so that it
corresponds to the relaxation below T$_{d}$ near 30\icm\ assigned
to the dynamics of polar nanoclusters. This relaxation is
responsible for the pronounced dielectric anomaly near and above
T$_{c}$. The relaxation does not completely vanish at low
temperatures, because our dielectric data at 1\,MHz and 8\,GHz
obtained on PST thin films of various B-site order show
permittivity between 300 and 400 at 20\K,\cite{Brinkman04} whereas
the phonon contribution to permittivity is only about 100 (see
Fig.~\ref{fig:PSTo_eps}). This is not surprising because a small
dielectric dispersion (constant dielectric losses) was observed in
many relaxors at low temperatures.\cite{Kamba04a,Buixaderas04}

The lowest-frequency phonon (marked TO$_1$) with the symmetry
F$_{\rm 1u}$ has the frequency near 45\icm{} at high temperatures.
It starts to harden below 550\K{} and it seems to split below
$T_{\rm c}$. For illustration, in Fig.~\ref{fig:PSTo_mod} we
fitted the temperature dependences of both components by the
Cochran law, but rather different extrapolated critical
temperatures (700 and 1000 K, respectively) were obtained. The
realistic value could be somewhere in between, but anyway, the
softening ceased near 600 K and above the Burns temperature of
about 700 K the expected hardening is not appreciable.

\begin{figure}[htbp]
  \begin{center}
    \includegraphics[width=\columnwidth]{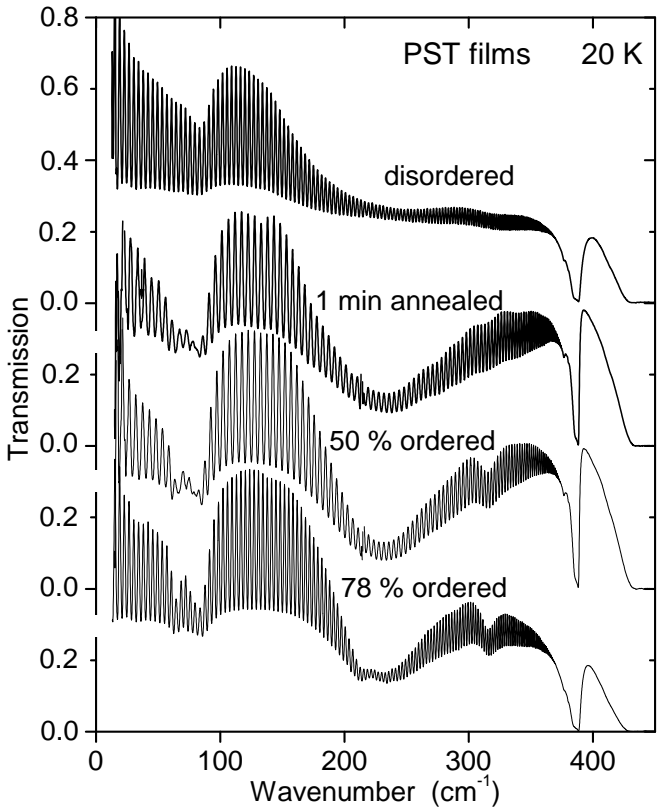}
  \end{center}
  \caption{Low-temperature IR transmission spectra of PST films
  with various B-site order.}
  \label{fig:PSTvar_Tr}
\end{figure}

Both components of the TO$_1$ SM have their counterparts also in Raman
spectra,\cite{Bismayer89}, but the assignment of the modes in Raman spectra is more
speculative. One mode near 56\icm\ dominates the Raman spectrum below 100\icm\ in the
paraelectric phase.\cite{Bismayer89} If we assume that it is Raman active due to the
B-site ordering, it should be of F$_{2g}$ symmetry, stemming from the Brillouin zone
boundary of simple cubic perovskite structure. Its frequency is by about 10\icm\ higher
than that of the F$_{1u}$ mode in our IR spectra (Fig.~\ref{fig:PSTo_mod}). At low
temperatures a triplet appears in Raman spectra with frequencies about 50, 60 (strong
peak) and 80\icm\ (at 100\K).\cite{Bismayer89} The mode near 80\icm\ corresponds to the
A$_{1}$ component of our TO$_{1}$ SM, which may activate in Raman spectra below T$_{c}$.
The mode near 50\icm\ could be assigned to the E component of the cubic F$_{2g}$ mode and
is seen only in Raman spectra, although it could be also (weakly) IR active. The mode
near 60\icm\ frequency could correspond to the E component of the TO$_{1}$ SM (see
Fig.~\ref{fig:PSTo_mod}) or to A$_{1}$ mode from the split F$_{2g}$ mode. The latter
explanation is more realistic because this mode is the strongest one in the Raman
spectra, and its strength decreases in IR spectra with increasing disorder at the B-sites
(see Fig.~\ref{fig:PSTvar_Tr}). This would mean that the mode near 30\icm\ should be
assigned to the E component of the polar SM (F$_{1u}$). This mode should remain IR active
up to 700\K\ due to the dielectric anisotropy in polar clusters\cite{Hlinka05b}, but it
is probably screened by the CM at high temperatures. This assignment sheds a new light on
softening of both modes near 80 and 60\icm. The higher-frequency mode softens and because
both modes are of the same A$_{1}$ symmetry so that they can bilinearly couple which
results also in partial softening of the lower-frequency mode due to the mode repulsion.

Let us discuss the question how the polar-phonon parameters in PST
are influenced by the different B-site ordering. The difference is
best remarkable at 20\K, where the phonon damping is minimized and
the phonons up to 400\icm{} can be resolved in our IR spectra.
Transmission spectra of four PST films with various order are
compared at 20\K{} in Fig.~\ref{fig:PSTvar_Tr}. One can see that
the TO$_2$ splitting (two minima near 230\icm) is appreciable only
in the highest ordered film. TO$_1$ splitting below 100\icm\ is
less and less pronounced with the decrease in ordering, but mostly
due to the increase in the linewidths.

Let us note that the film annealed at 850\celsius{} for 1$\;$min does not show any
ordering in XRD,\cite{Brinkman04} however remarkable difference between the disordered
and 1$\;$min ordered film is seen in our IR spectra. It shows that some short range
ordering occurs already in the 1$\;$min annealed film, but XRD is not sensitive enough to
see it.

Reaney \textit{et al.}\cite{Reaney94} correlated the intensity of
Sc-Ta stretching mode in PST seen at 315\icm{} with the degree of
B-site order. One can see this mode clearly in 50\% and 78\%
ordered films, in the other two samples only a~weak shoulder
appears (Fig.~\ref{fig:PSTvar_Tr}). Unfortunately, we cannot
establish any quantitative correlation between the intensity of
this mode and the degree of order. Namely, the spectra of 78\%
ordered and disordered films were measured on substrates which
were not polished on the bottom surface, while the other two
substrates were well polished on both surfaces. Matted bottom
surface causes parasitic diffuse scattering of the IR beam at
higher frequencies which is responsible for the lower value of the
transmission peak at 400\icm{} in the corresponding film spectra.
Fortunately, the quality of the rear substrate surface has no
influence on our spectra below 200\icm\ because the IR beam with
wavelength larger than 50$\;\mu$m does not feel its roughness.
\begin{figure}[htbp]
  \begin{center}
    \includegraphics[width=\columnwidth]{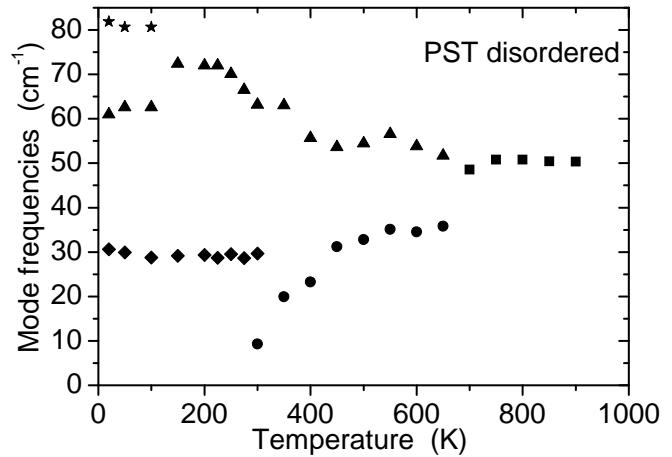}
  \end{center}
  \caption{Temperature dependences of the polar mode frequencies
  in disordered PST film.}
  \label{fig:PSTdis_mod}
\end{figure}

Temperature dependences of the mode frequencies in disordered PST
film are plotted in Fig.~\ref{fig:PSTdis_mod}. The mode near
60\icm\ appears only below 100\K{} due to the line broadening
compared to ordered film. Its lower strength than in ordered
sample gives evidence that it is the A$_{1}$ component of the
F$_{2g}$ Raman mode. The A$_{1}$ component of the TO$_1$ SM
exhibits behavior similar to that in the 78\% ordered film: it
softens on heating towards the Burns temperature which ceases
above 700\K. CM appears below 650\K{} and slows down below our
spectral range on cooling. The new heavily damped mode (probably
the E component of the F$_{1u}$ SM) appears below $T_{\rm c}$ at
30\icm{}. This mode was observed in all studied PST films as well
as in the PMN film\cite{Kamba05}. In Ref.\cite{Kamba05} we
suggested that the overdamping of the SM in the inelastic neutron
scattering spectra of PMN (so called "waterfall" effect) is only
an apparent effect. We suggested that the SM were underdamped (as
in IR spectra) but disappeared from the neutron spectra due to its
overlapping with the CM which approached the THz range near
$T_{\rm d}$. The SM and CM behavior in PMN and PST is therefore
very similar, but to our knowledge, PST was not studied by
inelastic neutron scattering so that no similar comparison is
possible.

In conclusion, our IR transmission spectra of variously ordered
PST thin films show qualitatively the same behavior: soft TO$_1$
phonon of F$_{1u}$ symmetry splits on cooling below the Burns
temperature. A$_{1}$ component hardens and follows the Cochran law
with extrapolated critical temperature slightly above the Burns
temperature. The low-frequency E component is clearly seen at low
temperatures, but it is probably active up to T$_{d}$, where it is
screened by the overdamped CM with similar frequency. CM stemming
from dynamics of polar clusters appears below the Burns
temperature and slows down to $T_{\rm c}$. Similar behavior was
observed also in the PMN relaxor,\cite{Kamba05} so it seems that
such a phonon and CM behavior is general for most perovskite
relaxor ferroelectrics.

\begin{acknowledgments}
We are grateful to J. Hlinka for useful discussions. This work was supported by the Grant
Agency of Academy of Sciences (projects Nos. A1010213 and AVOZ10100520), Grant Agency of
the Czech Republic (projects No. 202/04/0993) and Ministry of Education of the Czech
Republic (project COST OC 525.20/00).
\end{acknowledgments}

\end{document}